\documentclass[prl,showpacs,twocolumn]{revtex4-1}
\usepackage{graphicx}
\usepackage{color}

\begin{document}

\title{Quantum Thermal Transistor}
\author{Karl Joulain, J\'er\'emie Drevillon, Youn\`es Ezzahri and Jose Ordonez-Miranda}
\email{Corresponding author : karl.joulain@univ-poitiers.fr}
\affiliation{Insitut Pprime, CNRS, Universit\'e de Poitiers, ISAE-ENSMA, F-86962 Futuroscope Chasseneuil, France}

\date{\today}
\begin{abstract}
We demonstrate that a thermal transistor can be made up with a quantum system of three interacting subsystems, coupled to a thermal reservoir each. This thermal transistor is analogous to an electronic bipolar one with the ability to control the thermal currents at the collector and at the emitter with the imposed thermal current at the base. This is achieved by determining the heat fluxes by means of the strong-coupling formalism. For the case of three interacting spins, in which one of them is coupled to the other two, that are not directly coupled, it is shown that high amplification can be obtained in a wide range of energy parameters and temperatures. The proposed quantum transistor could, in principle, be used to develop devices such as a thermal modulator and a thermal amplifier in nanosystems.
\end{abstract}

\maketitle

Managing and harvesting wasted heat in energy processes is becoming a big issue due to the limited energy resources and to the constraints of global warming. Heat can be transported by fluids and radiation, as well as guided in good conductors or devices, such as heat pipes. However, there exists no device that can manage the switching or heat amplification, as is the case in electricity. 

In the last century, electricity management and its use for logical operations have been realized through the development of two components: the diode \cite{lashkaryov_investigations_1941} and the transistor \cite{bardeen_transistor_1998}. By analogy, one can, of course, envisage developing similar thermal devices that could make the thermal control easier. Thus, one of the goals of recent researches in thermal science has been focused on thermal rectifiers, i.e. components which exhibit an asymmetric flux when the temperatures at their ends are inverted. Thermal rectifiers have been designed for phononic \cite{terraneo_controlling_2002,li_thermal_2004,li_interface_2005,chang_solid-state_2006,hu_asymmetric_2006,yang_thermal_2007,hu_thermal_2009,pereira_sufficient_2011,zhang_thermal_2011,roberts_review_2011,garcia-garcia_thermal_2014}  and electronic \cite{roberts_review_2011,segal_single_2008} thermal transport, which has led to the conception and modeling of thermal transistors \cite{wang_thermal_2007,chung_lo_thermal_2008}. In the framework of thermal radiation, rectifiers have been the subject of numerous theoretical works, both in near field \cite{otey_thermal_2010-1,basu_near-field_2011,ben-abdallah_phase-change_2013} and far field \cite{van_zwol_emissivity_2012,ito_experimental_2014,nefzaoui_simple_2014,nefzaoui_radiative_2014,joulain_radiative_2015}. The most efficient of these devices have involved phase change materials, such as thermochrome \cite{huang_thermal_2013} materials like VO$_2$\cite{morin_oxides_1959,rini_photoinduced_2005}. This has led to the design of radiative thermal transistors based on phase change materials too\cite{ben-abdallah_near-field_2014,joulain_modulation_2015}.

The last two decades have also seen the emergence of individual quantum systems, such as classical atoms \cite{brune_quantum_1996,maunz_cavity_2004} or artificial ones, as is the case of quantum dots \cite{claudon_-chip_2009,dousse_ultrabright_2010}, which have been proposed to develop photon rectifiers \cite{yu_complete_2009,mascarenhas_quantum_2014,mascarenhas_quantum_2015}, transistors \cite{hwang_single-molecule_2009,astafiev_ultimate_2010} or even electrically controlled phonon transistors \cite{jiang_phonon_2015}. Moreover, given that quantum systems are always coupled to their environment, in particular to a thermal bath, the question of how heat is transferred through a set of quantum systems in interaction naturally arises \cite{manzano_quantum_2012,bermudez_controlling_2013,pumulo_non-equilibrium_2011}  and has led to several studies reporting thermal rectification \cite{scheibner_quantum_2008,pereira_symmetry_2009,werlang_optimal_2014,chen_thermal_2015}.

The goal of this Letter is to demonstrate that a thermal transistor can be achieved with a quantum system, made of 3 two level systems (TLS), which are equivalent to the three entries of a bipolar electronic transistor. It is shown that a thermal current imposed at the base can drive the currents at the two other entries of the system. More importantly, we also show these currents' perturbations imposed at the system entry can be amplified.

\begin{figure}
\begin{center}
\includegraphics[width=7cm]{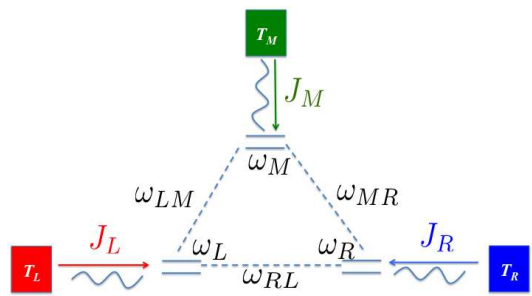}
\caption{Quantum system made up of 3 TLS coupled with each other and connected to a thermal bath.}
\label{system}
\end{center}
\end{figure}

The system under consideration consists of three TLS coupled with each other, each of them being connected to a thermal bath (Fig. \ref{system}). 
The three TLS are labeled with the letters $L$ (left), $M$ (medium), and $R$ (right), as well as the temperature of the thermal bath to which they are related.  We generalized the strong-coupling formalism developed by Werlang {\it et al.} \cite{werlang_optimal_2014} to the system considered here. Indeed, strong-coupling formalism is required to catch thermal effects, such as thermal rectification, that appear in a quantum system composed by two TLS \cite{werlang_optimal_2014}. Thermal transistor effects, which we want to emphasize in our study, are close to rectification effects and we therefore chose to follow the same approach. 
TLS, in our present case, are interacting spins that can be in the up state $\uparrow$ or in the down one $\downarrow$. The Hamiltonian of the system is (in $\hbar=1$ units) 
\begin{equation}
\label{ }
H_S=\sum_{P=L,M,R}\frac{\omega_P}{2}\sigma_z^P+\sum_{P,Q=L,M,R \ P\neq Q}\frac{\omega_{PQ}}{2}\sigma_z^P\sigma_z^{Q}
\end{equation}
where $\sigma_z^P$ ($P=L,M,R$) is the Pauli matrix $z$, whose eigenstates for the system $P$ are the states $\uparrow$ and $\downarrow$. $\omega_P$ denotes the energy difference between the two spin states, whereas $\omega_{PQ}$ stands for the interaction between the spin $P$ and the spin $Q$. $H_S$ eigenstates are given by the tensorial product of the individual TLS states, so that we have eight eigenstates labeled as $\left|1\right>=\left|\uparrow\uparrow\uparrow\right>$, $\left|2\right>=\left|\uparrow\uparrow\downarrow\right>$, $\left|3\right>=\left|\uparrow\downarrow\uparrow\right>$, $\left|4\right>=\left|\uparrow\downarrow\downarrow\right>$, $\left|5\right>=\left|\downarrow\uparrow\uparrow\right>$, $\left|6\right>=\left|\downarrow\uparrow\downarrow\right>$, $\left|7\right>=\left|\downarrow\downarrow\uparrow\right>$ and $\left|8\right>=\left|\downarrow\downarrow\downarrow\right>$. The coupling between the TLS and the thermal bath constituted of harmonic oscillators \cite{caldeira_quantum_1983} is based on the spin-boson model in the $x$ component
$H_{\rm TLS-bath}^P=\sigma_x^P\sum_k g_k(a_k^P a_k^{P\dag})$.
The three reservoirs $P$ have their Hamiltonians equal to
$H_{\rm bath}^P=\sum_k\omega_ka_k^{P\dag}a_k^P$.
This modeling implies that baths can flip one spin at a time. This means that there are 12 authorized transitions. The left bath ($L$) induces  the transitions $1\leftrightarrow5$, $2\leftrightarrow 6$, $3\leftrightarrow7$, and $4\leftrightarrow8$, the middle one ($M$) drives the transitions $1\leftrightarrow3$, $2\leftrightarrow 4$, $5\leftrightarrow7$, and $6\leftrightarrow8$. The right bath ($R$) triggers the transitions $1\leftrightarrow2$, $3\leftrightarrow 4$, $5\leftrightarrow6$, and $7\leftrightarrow8$. All other transitions flipping more than one spin are forbidden. 

The system state is described by a density matrix, which obeys a master equation. In the Born-Markov approximation, it reads
\begin{equation}
\label{master}
\frac{d\rho}{dt}=-i[H_s,\rho]+{\cal{L}}_L[\rho]+{\cal{L}}_M[\rho]+{\cal{L}}_R[\rho].
\end{equation}
As in \cite{werlang_optimal_2014,breuer_theory_2002}, the Lindbladians ${\cal{L}}_P[\rho]$ are written for an Ohmic bath according to classical textbooks \cite{leggett_dynamics_1987,breuer_theory_2002}, so that we take expression (4) of \cite{werlang_optimal_2014}.
We now consider a steady state situation. We define Tr$(\rho {\cal{L}}_P[\rho])=J_P$,  the heat current injected by the bath $J$ into the system.  Averaging the master equation we find $J_L+J_M+J_R=0$, in accordance with the energy conservation.

The master equation is a system of eight equations on the diagonal elements $\rho_{ii}$. If we introduce the net decaying rate from state $\left|i\right>$ to the state $\left|j\right>$ due to the coupling with bath $P$ with the help of Bose-Einstein distribution $n_\omega^P=(e^{\omega/T_P}-1)^{-1}$ (in $k_b=1$ units):
$\Gamma_{ij}^P=\omega_{ij}\left[\left(1+n_\omega^P\right)\rho_{ii}-n_\omega^P\rho_{jj}\right]=-\Gamma_{ji}^P$,
the master equation yields
\begin{eqnarray}
\dot{\rho}_{11} & = & 0 = \Gamma_{51}^L + \Gamma_{31}^M + \Gamma_{21}^R, \nonumber  \\
\dot{\rho}_{22} & = & 0 = \Gamma_{62}^L + \Gamma_{42}^M + \Gamma_{12}^R,\nonumber \\
\dot{\rho}_{33} & = & 0 = \Gamma_{73}^L + \Gamma_{13}^M + \Gamma_{43}^R, \nonumber \\
\dot{\rho}_{44} & = & 0 = \Gamma_{84}^L + \Gamma_{24}^M + \Gamma_{34}^R, \nonumber \\
\dot{\rho}_{55} & = & 0 = \Gamma_{15}^L + \Gamma_{75}^M + \Gamma_{65}^R,  \\
\dot{\rho}_{66} & = & 0 = \Gamma_{26}^L + \Gamma_{86}^M + \Gamma_{56}^R, \nonumber \\
\dot{\rho}_{77} & = & 0 = \Gamma_{37}^L + \Gamma_{57}^M + \Gamma_{87}^R, \nonumber \\
\dot{\rho}_{88} & = & 0 = \Gamma_{48}^L + \Gamma_{68}^M + \Gamma_{78}^R. \nonumber
\end{eqnarray}
These eight equations are not independent since their sum is 0. In order to solve the system for the $\rho_{ii}$, one adds the condition $Tr \rho=1$. Its resolution gives access to all state occupation probabilities as well as to the currents $J_J$.

Let us now explain what we call a thermal transistor effect, by making the analogy with an electronic one in which the current at the base controls the currents at the collector and at the emitter. A transistor effect is obtained when the collector and emitter currents can be modulated, switched and amplified by the current imposed at the base. The gain of the transistor is defined as the ratio of the current change at the collector or the emitter to the current variation applied at the base. Here, our goal is to show that it is similarly possible to control $J_L$ or $J_R$ by slightly changing $J_M$. We consider that the left and right TLS are both connected to thermal baths, whose respective temperatures $T_L$ and $T_R$ are fixed. The third bath at temperature $T_M$ controls the fluxes $J_L$ and $J_R$ with the help of a current $J_M$ injected into the system. Let us define the dynamical amplification factor $\alpha$:
\begin{equation}
\label{ }
\alpha_{L,R}=\frac{\partial J_{L,R}}{\partial J_M}.
\end{equation}
If a small change in $J_M$ makes a large change in $J_L$ or $J_R$, i.e. $|\alpha_{L,R}|>1$ then a thermal transistor effect will be observed in the same way as a large collector-current change is present by applying a small electrical current at the base of a bipolar transistor. 

We now focus on the conditions for which such a thermal transistor effect can be observed: a transistor will be characterized by the frequencies $\omega_P$, $\omega_{PQ}$ and the temperatures $T_L$ and $T_R$.  The last temperature $T_M$, that is taken here between  $T_L$ and $T_R$, controls the transistor properties and is related to the current $J_M$ through the current conservation condition. A fine parametric study of the system solutions for the density matrix is difficult, due to the large number of parameters. To reduce this number and to focus on the physics involved, we restrict our analysis to a simple case for which the two couplings  $\omega_{LM}=\omega_{MR}=\Delta$ whereas the last coupling $\omega_{RL}$ and the three TLS energies  are equal to 0. As shown below, this configuration provides a good transistor effect that can be interpreted with simple calculations. Note that the transistor effect disappears when the three couplings are equal (symmetric configuration), but it still occurs and can even be optimized if the three TLS energies are nonzero, but lower than $\Delta$ as discussed in the Supplemental Material \cite{SuppMat}. The operating temperature $T_L$ is taken so that $e^{-\Delta/T_L}\ll1$ ($T_L/\Delta\lesssim 0.25$) whereas $e^{-\Delta/T_R}\ll e^{-\Delta/T_L}$ ($T_R/\Delta\lesssim 0.0625$).

\begin{figure}
\begin{center}
\includegraphics[width=7cm]{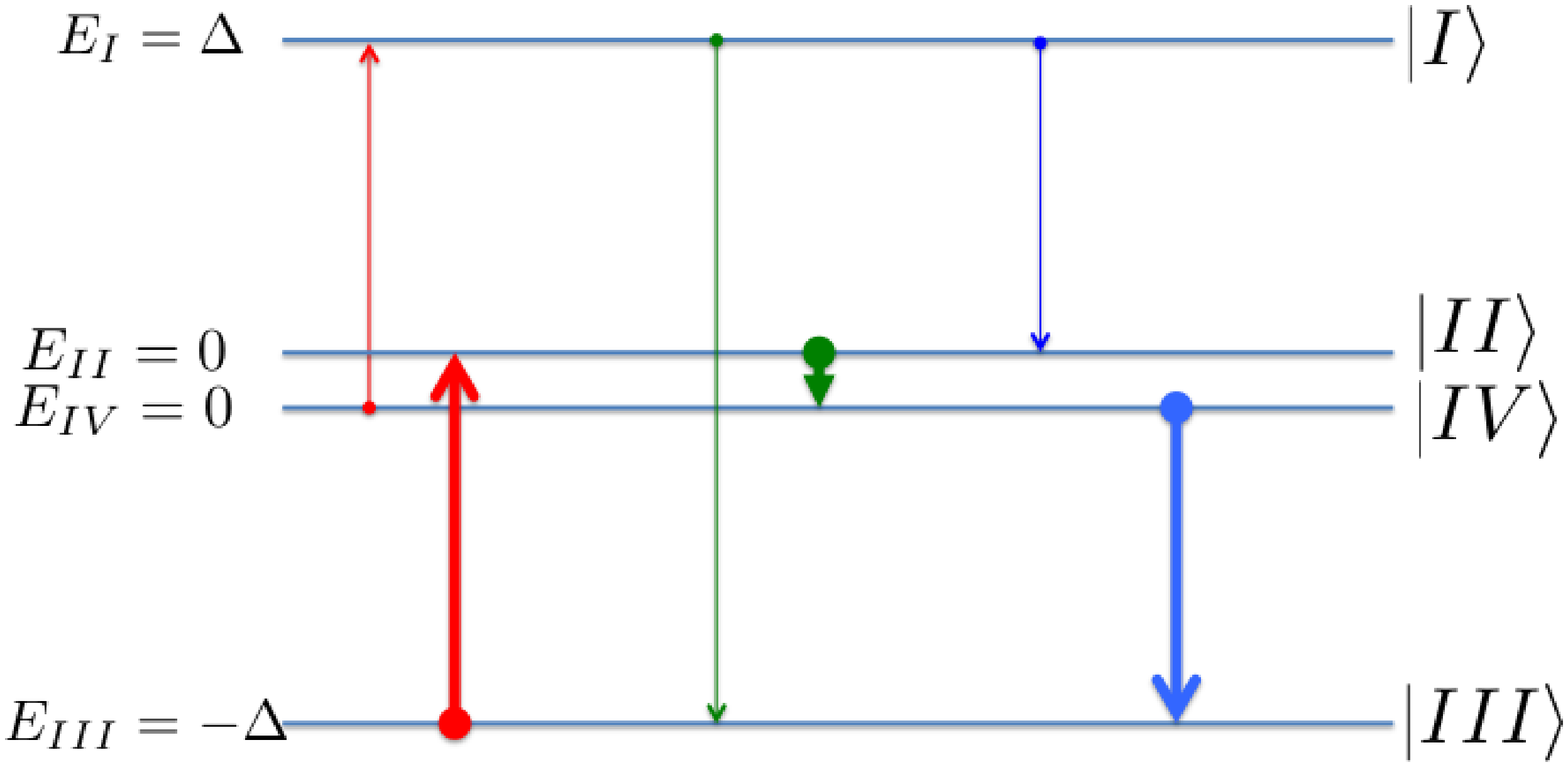}
\caption{Energy levels for $\omega_L=\omega_M=\omega_R=0$, $\omega_{RL}=0$, and $\omega_{LM}=\omega_{MR}=\Delta$. There are four states ($\left|I\right>$, $\left|II\right>$, $\left|III\right>$, and $\left|IV\right>$ but  three energy levels since $E_{II}=E_{IV}=0$. The arrows indicates the net decaying rate between the states due to bath $L$ (red), bath $M$ (green), and bath $R$ (blue)  for $T_L=0.2\Delta$, $T_R=0.02\Delta$, and $T_M=0.1\Delta$.}
\label{Energ_Levels}
\end{center}
\end{figure}

Under these conditions, the system states are degenerated two by two and there are only three energy levels (see Fig. \ref{Energ_Levels} and Supplemental Material \cite{SuppMat}). We rename the states $\left|1\right>$ and $\left|8\right>$ as $\left|I\right>$, $\left|2\right>$ and $\left|7\right>$ as $\left|II\right>$, $\left|3\right>$ and $\left|6\right>$ as $\left|III\right>$, and $\left|4\right>$ and $\left|5\right>$ as $\left|IV\right>$. We introduce the new density matrix elements, $\rho_I=\rho_{11}+\rho_{88}$, $\rho_{II}=\rho_{22}+\rho_{77}$, $\rho_{III}=\rho_{33}+\rho_{66}$, and $\rho_{IV}=\rho_{44}+\rho_{55}$. Introducing the net decaying rates between these states, the three currents simply read
\begin{eqnarray}
J_L & = & -\Delta\left[\Gamma^L_{I-IV}+\Gamma^L_{II-III}\right] \nonumber\\
J_M & = & -2\Delta\Gamma^M_{I-III}\\
J_R & = &   -\Delta\left[\Gamma^R_{I-II}+\Gamma^R_{IV-III}\right] \nonumber
\end{eqnarray}

\begin{figure}
\begin{center}
\includegraphics[width=7cm]{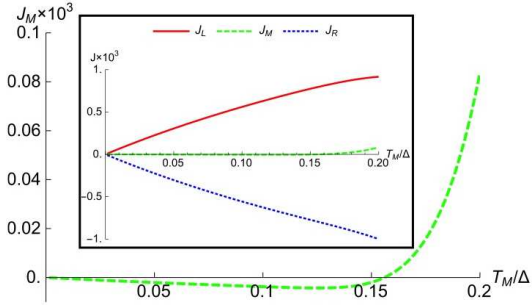}
\caption{Inset: thermal currents $J_L$, $J_M$, and $J_R$  versus $T_M$ for $\omega_L=\omega_M=\omega_R=0$, $\omega_{RL}=0$, $\omega_{LM}=\omega_{MR}=\Delta$, $T_L=0.2\Delta$, and $T_R=0.02\Delta$. Main figure: thermal current $J_M$ versus $T_M$. }
\label{three_currents}
\end{center}
\end{figure}

Transitions between the different states are illustrated in Fig. \ref{Energ_Levels}, for $T_L/\Delta=0.2$, $T_R/\Delta=0.02$, and $T_M/\Delta=0.1$. The arrow directions show the transition direction whereas its width depends on the decay time.  We see that energy exchanges are mainly dominated by the $III-II$ and $IV-III$ transitions. One therefore expects $J_R$ and $J_L$ to be larger than $J_M$. This is illustrated in Fig. \ref{three_currents}, where $J_L$, $J_M$, and $J_R$ are represented versus $T_M$, for $T_L/\Delta=0.2$ and $T_R/\Delta=0.02$. 
The two currents $J_L$ and $J_R$ increase linearly with temperature $T_M$, at low temperature, and become sublinear as $T_M$ approaches $T_L$. Note that over the whole range, as expected, $J_M$ remains lower than $J_L$ and $J_R$. Thus, $T_M$ will be controlled by changing slightly the current $J_M$: a tiny change of $J_M$ is therefore able to change significantly the values of $J_L$ and $J_R$. $J_L$ and $J_R$ can even be switched off when $J_M$ approaches 0 for small temperatures $T_M$,  so that the system exhibits the transistor switching property. Moreover, one sees that the $J_M$ slope remains larger than the ones of $J_L$ and $J_R$ over a large part of the temperature range. Given the definition of the amplification factor $\alpha$, the comparison of the thermal currents slope is the key element to see if amplification is present. 
\begin{figure}
\begin{center}
\includegraphics[width=7cm]{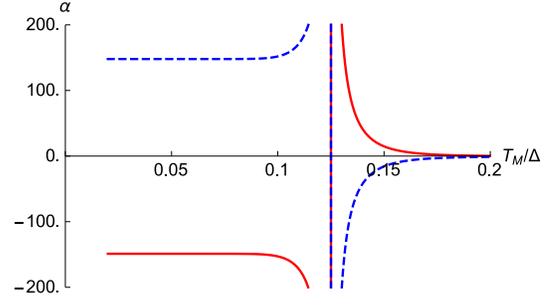}
\caption{Amplification factors $\alpha_L$ (red) and $\alpha_R$ (dashed blue) versus $T_M$ for $\omega_L=\omega_M=\omega_R=0$, $\omega_{RL}=0$, $\omega_{LM}=\omega_{MR}=\Delta$, $T_L=0.2\Delta$ and $T_R=0.02\Delta$.}
\label{alpha}
\end{center}
\end{figure}

In Fig. \ref{alpha}, we plot the two amplification coefficients $\alpha_L$ and $\alpha_R$ versus temperature $T_M$. We see that at low $T_M$, $\alpha$ remains much larger than 1. One also notes that $\alpha$ diverges for a certain value of the temperature for which $J_M$ has a minimum. This occurs for $T_M\simeq 0.1251\Delta$. In these conditions, an infinitely small change in $J_M$ makes a change in $J_L$ and $J_R$. As $T_M$ approaches $T_L$, the amplification factor drastically decreases to reach values below 1, i.e a regime where we cannot speak anymore of a transistor effect. Note also that, in between,  there exists a temperature where $J_M=0$. This is the temperature for which the bath $M$ is at thermal equilibrium with the system since it does not inject any thermal current in it. At this temperature ($T_M\simeq 0.156\Delta$), $J_L=-J_R=7.97\times10^{-4}$. Amplification still occurs since $\alpha_L=8.88$ and $\alpha_R=-9.88$.

All these observations can be explained by examining carefully the populations and currents expressions (see Supplemental Materials \cite{SuppMat} for details). In the present case, if we limit the calculation to first order of approximations on $e^{-\Delta/T_L}$ and $e^{-\Delta/T_M}$,  one can roughly estimate the populations by
\begin{eqnarray}
\rho_I & \simeq & \frac{e^{-2\Delta/T_M}}{2}+\frac{T_M}{4\Delta+8T_M}e^{-2\Delta/T_L} \label{rho1},\\
\rho_{II} & \simeq & \frac{\Delta+T_M}{\Delta+2T_M}e^{-\Delta/T_L}\label{rho2},\\
\rho_{III} & \simeq & 1-e^{-\Delta/T_L}\label{rho3},\\
\rho_{IV} & \simeq & \frac{T_M}{\Delta+2T_M}e^{-\Delta/T_L}\label{rho4}.
\end{eqnarray}
$\rho_{III}$ remains very close to 1 and $\rho_{II}$ to 10$^{-2}$. $\rho_{I}$ and $\rho_{IV}$
are much lower but change by 1 to 2 orders of magnitude with temperature. Note that the sum of (\ref{rho1}-\ref{rho4}) is not 1, but  the error on $\rho$'s is less than 1\% over the temperature range.

We now explicitly correlate the three thermal currents with temperature.
\begin{eqnarray}
J_L & \simeq & -J_R \simeq \frac{\Delta^2 T_M e^{-\Delta/T_L}}{\Delta+2T_M}, \\
J_M & \simeq &\Delta^2\left[-\frac{T_M}{\Delta+2T_M}e^{-2\Delta/T_L}+2 e^{-2\Delta/T_M}\right].
\end{eqnarray}

Note here again that the sum of the three currents is not zero despite of the fact that expressions remain close to the exact solution. The largest error is on $J_L$ and $J_R$ around $T_M=T_L$ where it reaches 5\%. These formula retrieve the linear dependence of the thermal currents for small values of $T_M$. One also notes that, when we compare with (\ref{rho4}), $J_L$ and $J_R$ are driven by $\rho_{IV}$, i.e., the state population at the intermediate energy ($E_{IV}=0$). Looking at the authorized transitions, one expects $J_M$ to be driven by the population of the most energetic state, i.e., $\rho_I$. The main difference between $\rho_{IV}$ and $\rho_I$ is the temperature dependence, which is linear in one case and exponential ($e^{-2\Delta/T}$) in the other case. The result is that even when $T_M$ is close to $T_L$, $\rho_I$ remains low. Therefore, $J_M$ keeps low values in the whole temperature range due to the low values of $\rho_I$.
If we look more carefully at $J_M$, one notices that it is the sum of two terms. The first one is roughly linear on $T_M$. It is similar to the one that appears in $\rho_{IV}$. $J_M$ depends on the population of state $IV$, which also influences the population of state $I$ with the transition $IV-I$. The increase of $\rho_{IV}$ with $T_M$ makes easier the $IV-I$ transition, and raises $\rho_I$. This increases the decaying of state $I$ through the $I-III$ transition. This term is negative and decreases as $T_M$ increases. This can be seen as a negative differential resistance since a decreasing of $J_M$ (cooling in $M$) corresponds to an increase of the temperature $T_M$. In this temperature range, it can easily be shown that the amplification factor $|\alpha_L|\approx|\alpha_R|\approx e^{\Delta/T_L}$. 
A second term in $J_M$, is the classical $e^{-\Delta/T_M}$ Boltzmann factor term, which makes the population of state $I$ increase with $T_M$. $J_M$ is a tradeoff between these two terms. At low temperature, the linear term is predominant. As $T_M$ increases, the term $e^{-\Delta/T_M}$ takes over. As a consequence, there is a point where $\rho_I$ increasing reverses the $I-IV$ transition, so that the $I-III$ transition competes with both $I-IV$ and $I-II$ transitions. $I-III$ is then reversed. With these two terms competing, there is a temperature for which $J_M$ reaches a minimum and a second temperature where $J_M=0$, as already described. 

One can summarize the conditions needed for the system to undergo a thermal transistor effect. Two baths (here $L$ and $R$) induce transitions between two highly separated states with an intermediate energy level, whereas the third one ($M$) makes only a transition between the two extremes. This will first make $J_M$ much smaller than $J_L$ and $J_R$, and second, it will set a competition between a direct decay of the highest level to the ground level and a decay via the intermediate one. This competition between the two terms makes the thermal dependance of $J_M$ with $T_M$ slow enough to obtain a high amplification. 

Finally, one can wonder what kind of real system could make such a thermal quantum transistor. A simple TLS related to a bath could be, for example, a quantum dot with a single bound state, embedded in a material at temperature $T$. The TLS is the quantum dot and the bath is the material at temperature $T$. The transistor proposed in this Letter could be made of three quantum dots each of them embedded in a nanoparticle. The three nanoparticles could be deposited on a substrate and the distance between them adjusted in order that the coupling energy between quantum dots reaches the desired value $\Delta$. The nanoparticles temperature could be controlled by electrical means.


In conclusion, we have shown that it is possible to make a thermal transistor with three coupled TLS linked to three different thermal baths. One TLS is coupled to the two others, whereas these last are not directly coupled. The TLS related to the two others plays the same role as the base in a bipolar transistor, while the two other TLS can be seen as the emitter and the collector. We found a temperature regime where a thermal current variation imposed at the base generates an amplified variation at the emitter and the collector. This regime is typically such that temperature corresponds to an energy one order of magnitude smaller  than the coupling energy between the TLS. This means that a transistor effect will be observed at ambient temperature for a coupling between the TLS with a typical frequency in the visible. With this kind of thermal transistor one can expect to modulate or amplify thermal fluxes in nanostructures made up of elementary quantum objects.

\begin{acknowledgments}
This work pertains to the French Government Program ''Investissement d'avenir'' (LABEX INTERACTIFS, ANR-11-LABX-0017-01). K.J. thanks Alexia Auff\`eves for fruitful and inspiring discussions.
\end{acknowledgments}



\end{document}